\documentclass[sigconf,nonacm,natbib=false]{acmart}

\author{Keno Hassler}
\authornote{Both authors contributed equally to this research.}
\orcid{0009-0008-9115-2769}
\affiliation{%
  \institution{CISPA Helmholtz Center for Information Security}
  \city{Saarbrücken}
  \country{Germany}}
\email{keno.hassler@cispa.de}
\author{Philipp Görz}
\authornotemark[1]
\orcid{0009-0001-0501-1249}
\affiliation{%
  \institution{CISPA Helmholtz Center for Information Security}
  \city{Saarbrücken}
  \country{Germany}}
\email{research@philipp-goerz.com}
\author{Stephan Lipp}
\orcid{0009-0008-4254-9145}
\affiliation{%
  \institution{Technical University of Munich}
  \city{Munich}
  \country{Germany}}
\email{stephan.lipp@tum.de}

\usepackage{amsmath,amsfonts,amsthm}
\usepackage{algorithm,algorithmic}
\usepackage{xspace}
\usepackage{graphicx}
\usepackage{textcomp}
\usepackage[svgnames]{xcolor}
\usepackage{listings}
\usepackage{hyperref}
\usepackage{pifont}
\usepackage{enumitem}
\usepackage{url}
\usepackage{multirow}
\usepackage{subcaption}
\usepackage{fontawesome5}
\usepackage[binary-units]{siunitx}

\usepackage{tabularx}
\usepackage{booktabs}

\usepackage{tikz}
\usepackage{tcolorbox}
\usepackage{threeparttable}
\usepackage{supertabular}
\usepackage{acro}
\usepackage{multicol}

\DeclareAcronym{asan}{
	short={ASAN},
	long={Address Sanitizer},
	cite={serebryany_addresssanitizer_2012}
}
\DeclareAcronym{sast}{
	short={SAST},
	long={Static Application Security Testing},
}
\DeclareAcronym{dast}{
	short={DAST},
	long={Dynamic Application Security Testing},
}
\DeclareAcronym{cwe}{
	short={CWE},
	long={Common Weakness Enumeration},
}
\DeclareAcronym{cve}{
	short={CVE},
	long={Common Vulnerabilities and Exposures},
}
\DeclareAcronym{plt}{
	short={PLT},
	long={Procedure Linkage Table},
}
\DeclareAcronym{ast}{
	short={AST},
	long={Abstract Syntax Tree},
}
\DeclareAcronym{sql}{
	short={SQL},
	long={Structured Query Language}
}
\DeclareAcronym{hmac}{
	short={HMAC},
	long={hash-based message authentication code}
}
\DeclareAcronym{sarif}{
	short={SARIF},
	long={Static Analysis Results Interchange Format},
	cite={fanning2018static},
}
\DeclareAcronym{put}{
	short={PUT},
	long={program under test},
}

\def\BibTeX{{\rm B\kern-.05em{\sc i\kern-.025em b}\kern-.08em
    T\kern-.1667em\lower.7ex\hbox{E}\kern-.125emX}}

\newtheorem*{theorem*}{Theorem}

\definecolor{mycolor}{rgb}{0.122, 0.435, 0.698}%
\definecolor{gray1}{gray}{0.3}

\definecolor{darkgreen}{rgb}{0.0, 0.5, 0.0}
\definecolor{darkred}{rgb}{0.82, 0.1, 0.26}
\newcommand{\cmark}{\textcolor{darkgreen}{\ding{51}}}%
\newcommand{\xmark}{\textcolor{darkred}{\ding{55}}\ }%

\lstdefinestyle{tango}{
    captionpos=b,
    frame=tb,
    breaklines=true,
	flexiblecolumns=true,
	showstringspaces=false,
	keepspaces=true,
    tabsize=4,
    basicstyle=\ttfamily\bfseries\footnotesize,
    identifierstyle=\mdseries,
    keywordstyle=\bfseries\color{blue!70!black},
    stringstyle=\mdseries\itshape\color{lime!60!black},
    commentstyle=\mdseries\color{orange!80!black},
    numbers=left,
    numberstyle=\ttfamily\tiny,
}
\newcommand{\cparagraph}[2][.35\baselineskip]{\par\noindent\textbf{#2.}\xspace}

\DeclareSIUnit{\cent}{\text{¢}}

\newcommand{\magma}{Magma\xspace}
\newcommand{\cgc}{CGC\xspace}

\newcommand{\abbrev}[1]{\textsl{#1}\xspace}

\newcommand{\etal}[1]{\mbox{#1}~\abbrev{et~al.}}

\newcommand{\tool}[1]{\textsc{\mbox{#1}}\xspace}

\newcommand{\clangsa}{\tool{Clang~SA}}
\newcommand{\codeql}{\tool{CodeQL}}
\newcommand{\flawfinder}{\tool{Flawfinder}}
\newcommand{\infer}{\tool{Infer}}
\newcommand{\semgrep}{\tool{SemGrep}}

\newcommand{\afl}{\tool{AFL}}
\newcommand{\aflfast}{\tool{AFLFast}}
\newcommand{\aflpp}{\tool{AFL{++}}}
\newcommand{\angora}{\tool{Angora}}
\newcommand{\entropic}{\tool{Entropic}}
\newcommand{\fairfuzz}{\tool{FairFuzz}}
\newcommand{\honggfuzz}{\tool{Honggfuzz}}

\newcommand{\libfuzzer}{\tool{LibFuzzer}}
\newcommand{\moptafl}{\tool{MOpt-AFL}}
\newcommand{\parmesan}{\tool{ParmeSan}}
\newcommand{\symccafl}{\tool{SymCC-AFL}}

\newcommand{\aflppthree}{\aflpp~(3.00a)}
\newcommand{\libafl}{\tool{libafl\_libfuzzer}}

\newcommand{\subject}[1]{\mbox{#1}\xspace}

\newcommand{\github}{\subject{GitHub}}

\newcommand{\libpng}{\subject{Libpng}}

\newcommand{\libxml}{\subject{Libxml2}}

\newcommand{\openssl}{\subject{OpenSSL}}
\newcommand{\php}{\subject{PHP}}
\newcommand{\poppler}{\subject{Poppler}}

\newcommand{\sqlite}{\subject{SQLite3}}

\newcommand{\numFuzzers}{13\xspace}
\newcommand{\numAnalyzers}{five\xspace}

\usetikzlibrary{external}
\input{generated/data.tex}

\usepackage[
  style=acmnumeric,
  doi=false,
  isbn=false,
  url=false,
]{biblatex}
\AtEveryBibitem{\clearlist{location}}
\AtEveryBibitem{\clearlist{publisher}}
\AtEveryBibitem{\clearfield{month}}
\AtEveryBibitem{\clearfield{day}}
\AtEveryBibitem{\clearfield{urlyear}}

\begin{document}
\title{A Comparative Study of Fuzzers and Static Analysis Tools for Finding Memory Unsafety in C and C++}

\begin{abstract}

Over 70\% of security vulnerabilities in critical software systems today result from memory safety violations.
To address this challenge, fuzzing and static analysis are widely used automated methods to discover such vulnerabilities.
Fuzzing generates random program inputs to identify faults at runtime, while static analysis reasons about the code to detect potential vulnerabilities.
Although these techniques share a common goal, they take fundamentally different approaches and have evolved largely independently.

In this paper, we present an empirical analysis of \numAnalyzers static analyzers and \numFuzzers fuzzers, applied to over 100 known security vulnerabilities in C/C++ programs.
We measure the detection rate for each tool and vulnerability to evaluate how the approaches differ and complement each other.
We find that fuzzers discover a very similar set of bugs, while static analyzers report more diverse sets,
and identify clear leaders for each group.
Comparing the union of all fuzzers with that of all static analyzers, we observe they are nearly disjoint.
In a second step, we manually validate the report-to-bug mapping we developed for the evaluation and discuss more qualitative aspects of limitations, usability, and integration into the development process.
We examine how widely these bug finding tools are used in critical open-source projects.

We advise developers on choosing tools to harden their software and identify barriers to adoption as well as future research opportunities.
\end{abstract}

\setcopyright{rightsretained}

\maketitle
\section{Introduction}
In recent years, more and more security vulnerabilities have been reported in critical software systems. 
Many of these vulnerabilities are related to memory unsafety. 
For instance, a third of \emph{all} recorded security vulnerabilities (tracked as \acsp*{cve}) and seven of the 20 most frequently reported bug classes (tracked as \acsp*{cwe}) are due to memory safety violations.
More than 70\% of reported vulnerabilities in the Chrome browser~\cite{taylor2022usefreedom} as well as in the iOS, macOS \cite{kehrer2019memory}, and Android~\cite{vanderstoep2019queue} operating systems~\cite{gaynor2019introduction} are related to memory unsafety. 
In fact, 72\% of zero-day vulnerabilities in the last 12 years are due to memory safety violations enabled by memory-unsafe languages~\cite{google20250day}.

To scale bug finding beyond manual code audits, developers have turned to automated vulnerability discovery tools based on techniques such as static analysis and fuzz testing (in short \emph{fuzzing}). 
For instance, Google considers fuzzing their first line of defense~\cite{taylor2021update}, while Meta reports that static analysis accounts for 70\% of vulnerabilities discovered internally~\cite{rodriguez2022how}.
Illustrating the popularity of fuzzing, the OSS-Fuzz project~\cite{serebryany2017ossfuzz} continuously tests more than 1{,}000 open source software (OSS) projects, including many popular and security-critical projects. 
Regarding the popularity of static analysis, \etal{Beller}~\cite{beller2016analyzing} found in 2016 that approximately half of the 122 surveyed open-source projects employ static analysis. 

Fuzzing detects bugs by provoking unexpected program states, while static analysis examines the program's source code. 
In a \emph{fuzzing setup}, a fuzzer passes automatically generated inputs to the program under test via a \emph{harness}---a fuzzer-specific entry point that is usually written manually.
In a \emph{static analysis setup}, only the static analyzer is required, though in some cases the program needs to be compiled. 
Unlike fuzzers, static analyzers are known to report false positives, i.e., benign code falsely flagged as buggy. 
But how do fuzzing and static analysis compare in their ability and effectiveness at detecting real bugs?

Fuzzing and static analysis share the fundamental goal of finding bugs in programs, yet there is little research comparing these two approaches systematically. 
This can be explained by two factors: first, these tools are developed by largely separate communities; second, fuzzing and static analysis tools require fundamentally different project integrations, essentially requiring extra work to integrate the tools. 
These differences make a systematic comparison of both approaches particularly interesting. 
Is it worth the effort to employ both approaches in parallel, and can the approaches benefit from learning from each other, bringing both research communities closer together?

In this paper, we adopt the perspective of an open-source C/C++ project maintainer who wants to use a bug finding tool to discover memory safety vulnerabilities.
Where should they focus their resources?
Currently, there are no clear guidelines for choosing one approach over the other. 
There is also no systematic analysis of their relative strengths and weaknesses. 
To develop such a systematic analysis, we define clear selection criteria and evaluate over 100 \acsp*{cve} across seven open-source projects (using \magma \cite{hazimeh2020magma}), employing \numAnalyzers static analyzers and \numFuzzers different fuzzers.
Concretely, we

\begin{itemize}[leftmargin=1em, nosep]
\item quantitatively compare static analysis tools and fuzzers from an end-to-end perspective (the number of bugs detected), exploring their strengths and weaknesses,
\item manually validate our metrics, evaluate the qualitative aspects of limitations, usability and resource usage and investigate how tools are used in the wild, and
\item derive recommendations for developers on how to add security tooling into their workflows.
\end{itemize}

We find that each category of tools has a clear winner, but both find almost disjoint sets of bugs. After manual analysis, a majority of the bugs in our dataset is not detected by any tool.

\cparagraph{Outline}
We start with an overview of the relevant background and related work (\autoref{sec:background}).
After defining the scope of our study (\autoref{sec:scope}), we conduct our analysis in two parts. 
The first part is an \emph{empirical evaluation} (\autoref{sec:quant-eval}) of freely available fuzzers and static analysis tools focusing on quantitative aspects of comparison.
In the second part, we conduct a \emph{qualitative analysis} (\autoref{sec:qual}), based on the empirical part.
We discuss the implications of our analysis for project maintainers and research communities in \autoref{sec:discussion} and conclude the paper in \autoref{sec:conclusion}.

\cparagraph{Data Availability}
To make our findings reproducible and foster further research in this direction, we make our evaluation scripts and data openly available at \textcolor{blue}{\url{https://anonymous.4open.science/r/fuzzing-vs-static-artifact-8973/}}.

\section{Background}
\label{sec:background}

In this section, we explain the types of vulnerabilities we are interested in, before briefly summarizing the state-of-the-art in bug finding tools and the benchmarks commonly used to evaluate them.

\subsection{Memory Unsafety in C/C++}
\label{sec:background-mem-unsafety}

In this work, we focus on one of the most prevalent classes of security vulnerabilities: memory unsafety and related types of vulnerabilities prevalent in C/C++ programs.
As discussed before, more than 70\% of the reported vulnerabilities in Chrome, iOS, macOS, and Android~\cite{taylor2022usefreedom, kehrer2019memory, vanderstoep2019queue} and 72\% of zero-days are due to memory safety violations~\cite{google20250day}.
Software systems can be subject to many types of bugs and vulnerabilities.
However, memory safety violations are (almost) unique to C/C++.
While many modern languages provide memory safety by design, in C/C++ it is primarily the developer's responsibility to uphold the safety contract.
In exchange, C/C++ offers more low-level control and (in absence of runtime checks and garbage collection) a high performance. %

To be specific, we exclude bugs that are generally not security-relevant, such as unused variables or violations of coding conventions reported by a linter.
Furthermore, bugs exclusive to other languages are not relevant for our analysis.
Lastly, security-critical bugs that are not related to memory safety violations, such as side-channel attacks or information leaks, are also considered out of scope.
This means we are evaluating the bug finding tools from a \emph{security} perspective (focusing on memory unsafety in C/C++): while we touch upon this aspect in \autoref{sec:qual}, we do not consider bug finding primarily as a tool to improve software \emph{quality}.

\subsection{Bug Finding: Static vs. Dynamic Methods}
We investigate two prevalent automated bug finding approaches: fuzzing and static analysis.
\cparagraph{Static Analysis}
\ac{sast} refers to a collection of techniques that examine source code without actually executing the code.
Notably, this always requires some degree of (over-)approximation,
with the trade-off being a varying number of false positive reports that flag a correctly implemented piece of code as defective.
At the coarse end of the spectrum lies \emph{syntactic analysis}, which is fundamentally pattern matching on the syntax level with known or suspicious antipatterns. This approach is easy to implement and inexpensive to run.
A more sophisticated approach is \emph{semantic analysis}, where checks are run on the properties of the program's \ac{ast}.
At the other end, there is \emph{symbolic execution}, a computationally expensive technique that simulates program paths by reasoning about symbolic states.

\cparagraph{Fuzzing}
Fuzzing~\cite{miller1990empirical} is a dynamic testing technique.
The \ac{put} is executed on automatically generated, concrete test inputs.
If an illegal state is reached, the fuzzer is notified that it has found a bug and saves the respective input to disk.
While it is left to a human analyst to find the root cause of this bug, all reports caused a crash during execution, indicating the presence of a bug.
Detection of illegal states is performed by sanitizers (cf. \etal{Song}~\cite{song2019sok}) that typically under-approximate, and thus, do not typically produce false positive reports.
Since the advent of AFL~\cite{zalewski2017american}, greybox fuzzers that use program coverage as an inexpensive feedback loop have become state-of-the-art.
The now-deprecated AFL has found large numbers of bugs itself~\cite{zalewski2017american}, and its descendant \aflpp~\cite{fioraldi2020afl}, which incorporates more recent research ideas, continues on that path.

\subsection{Related Work}

Both static analysis and fuzzing are active research domains, this includes extensive research evaluating fuzzers or static analyzers.
However, to the best of our knowledge, no cross-domain comparison has been attempted so far.
In the following, we give an overview of existing work.

\cparagraph{Static Analysis}
\etal{Sadowski}~\cite{sadowski2018lessons} report on their experiences with the development of static analysis tools at Google.
They suggest that tool authors should focus on the real needs of developers so that the tool can be integrated directly into the development workflow to find bugs early.
They also suggest that tools should be made open-source so that they can better keep pace with future bug finding challenges.
Obviously, this also requires constant evaluation of new and established static analyzers to see how well they actually find (security) bugs.

\etal{Johnson}~\cite{johnson2013why} find in a small survey that a lack of information in warning messages given to developers inhibits more wide-spread adoption.
\etal{Christakis}~\cite{christakis2016what} confirm this in a larger study, adding that developers expect a false-positive rate below 20\% and the ability to configure the tool.
A data scrape on StackOverflow by \etal{Imtiaz}~\cite{imtiaz2019challenges} supports these results, calling false positives a ``dominant issue'' for the adoption of static analysis tools.
Consequently, many research papers~\cite{kremenek2003zranking,livshits2005finding,kim2007which,ruthruff2008predicting,ayewah2008using,vassallo2018context,muske2018repositioning} have attempted to reduce the number of false positive warnings.
The comparative evaluation of static analyzers is a particularly active field of research.
Numerous papers evaluate the bug finding capabilities of analyzers on C/C++ source code across different synthetic and real-world benchmarks (e.g.,~\cite{zitser2004testing,zheng2006value,chatzieleftheriou2011testdriving,thung2012what,goseva-popstojanova2015capability,habib2018how,aloraini2019empirical,dabruzzopereira2020use,kaur2020comparative}).
Closely related to our work, \etal{Lipp}~\cite{lipp2022empirical} evaluate \ac{sast} tools on memory vulnerabilities, finding that the majority of bugs in the benchmark remain undetected by state-of-the-art tools.
In this study, we go a step further and compare the bug finding capabilities of \ac{sast} tools \emph{with those of fuzzers}, aiming to find differences and potential benefits from combining these tools.

\cparagraph{Fuzzing}
Fuzzing has been deployed in large-scale evaluations, notably in Google's OSS-Fuzz~\cite{serebryany2017ossfuzz}, an open-source software fuzzing service that continuously tests hundreds of popular projects.
Ding \emph{et al.}~\cite{ding2021empirical} analyze more than twenty thousand bugs, including thousands of severe bugs, found via OSS-Fuzz, demonstrating its real-world impact.
Furthermore, they underline that effective fuzzing is an iterative process that requires fixing detected bugs, allowing the fuzzing campaign to progress.
Google's Fuzzbench~\cite{metzman2021fuzzbench} framework offers a free benchmark service for fuzzer developers running on Google infrastructure.
The \magma benchmark~\cite{hazimeh2020magma} hand-picks a set of real-world bugs and provides perfect oracles for them, aiming to remove inaccuracies in comparisons.
\etal{Klees}~\cite{klees2018evaluating} and \etal{Schloegel}~\cite{schloegel2024sok} give recommendations on reproducible scientific evaluations for fuzzers.
\etal{Böhme} have shown that code coverage is strongly correlated with the number of bugs found~\cite{bohme2022reliability}, but discovering new bugs becomes exponentially harder~\cite{bohme2020fuzzing}.

\section{Scope of Comparative Study}
\label{sec:scope}

Our research is motivated by the common situation where a maintainer of an open-source C/C++ project decides to add a bug finding tool to discover memory safety issues in their project.
However, to assess the performance of a tool for this task systematically, we first need to identify a suitable benchmark.

\subsection{Benchmark Selection Criteria}
In order to make a fair comparison between fuzzers and \ac{sast} tools, we need a dataset with known, labeled vulnerabilities, including the type of each bug.
We require that the bugs are security-relevant and found in C or C++ programs.
For generalizability, the bugs should capture different real-world application domains, different input structures, and a variety of operations and transformations.
Finally, and crucially, we want to avoid using one tool group's output as ``ground truth'' for the other, as that would inherently place an upper bound on the other group's performance.
These requirements, together with the aim for an unbiased dataset, make it difficult to find a suitable set of bugs to use as a benchmark.

\cparagraph{Magma}
The \magma benchmark (v1.1) \cite{hazimeh2020magma} consists of 112 bugs in seven open-source programs that range from image parsers and a cryptographic library to database and website engines.
The (publicly known) bugs were re-inserted into a newer version of the respective codebase.
Most importantly, these bugs were chosen \emph{irrespective} of the technique used to find them.%
\footnote{
  Regarding the collation of bugs, the \magma authors write:
  ``No specific set of criteria was imposed on the bug selection process.
  However, throughout our porting efforts, we often prioritized more recent
  bug reports, since they correlate most closely to the latest code base, and
  are thus more likely to remain valid. Additionally, reports marked
  `critical' were also given a higher priority than others.''~\cite{hexhive2020frequently}
}
Thus, \magma fulfills our selection criteria.
Note that \magma necessarily reflects biases present in real-world vulnerability discovery since it is derived from previously reported vulnerabilities potentially detected using bug finding tools.
Such biases are unavoidable in any real-world benchmarks and cannot be corrected, as the distribution of undiscovered vulnerabilities in the wild is unknowable.
The \magma benchmark includes \emph{canaries}, a mechanism to detect if a specific bug is triggered during runtime.
For evaluating static analysis tools, there is metadata (a \acs{cve} ID) for each bug that maps it to a vulnerability class (\acs{cwe}).

\cparagraph{\cgc}
However, to make our evaluation as comprehensive as possible, we also consider other benchmarks.
Unfortunately, benchmarks with synthetic bugs are not as well-suited for our evaluation, as the bug types are unknown, it is unclear if the bugs violate a security goal, or the artificial programs are far from realistic.
For example, the Juliet benchmark~\cite{nist_2017_juliet} is a popular benchmark for static analysis tools.
In it, most (bug-containing) samples simply crash on execution or do not take any input, making them unsuitable for evaluation against dynamic tools like fuzzers.
From the fuzzing community, there is a bug-based benchmarking mode in FuzzBench~\cite{metzman2021fuzzbench}, but it is lacking any metadata about the bugs.
The only other benchmark that satisfies our necessary selection criteria is the set of DARPA Cyber Grand Challenge (CGC)~\cite{novafacing_cgc-challenges} binaries.
Therefore, to expand our evaluation, we also evaluate a subset of the tools on the \cgc binaries.
This benchmark consists of custom-written, small programs containing exploitable vulnerabilities, however, they are thus less realistic than \magma.
There are no canaries, but vulnerability descriptions that include \acsp{cwe} and the specific bug location.
We reflect more on the availability of bug-based benchmarks for \ac{sast} tools in \autoref{sec:lessons-learned}.

\subsection{Tool Selection Criteria}
To evaluate the discovery of memory safety vulnerabilities in C/C++ programs, we define selection and exclusion criteria for the tools in our evaluation.
This may limit the generalizability of our work, but is necessary to make this wide-scale comparison possible given the sheer amount of available tools.

In terms of selection criteria, we are interested in widely used or actively maintained tools~\cite{static_analysis_c, magma_survival_report} that can find memory safety violations in C/C++ programs, are open-source (or at least freely available), and cover a wide variety of approaches.
We exclude tools that report code quality issues rather than security-related issues, and tools that only have experimental support for memory safety issues.
Furthermore, we exclude tools that require substantial manual input from an expert (like grammar specifications for smart fuzzers or function contracts for model checkers), that are outdated or lack sufficient documentation for a reasonable setup, or that are incompatible with our experimental infrastructure (e.g., operating system and hardware).

\cparagraph{Static analysis tools}
We distinguish five syntactic and semantic analyzers that satisfied our criteria:
\begin{itemize}[leftmargin=1em, nosep]
  \item \emph{Syntactic (source code)}. \flawfinder (v2.0.19)~\cite{wheeler2001flawfinder} encodes well-known vulnerability patterns as regular expressions that can be efficiently and most generally matched on the source code for vulnerability detection. %
  \semgrep (v1.24.0)~\cite{semgrep} additionally takes the semantics of a programming language into account, thereby reducing the number of false positives while keeping the analysis independent of the build system. Moreover, \semgrep decouples its rule set~\cite{semgrep_rules} from the analysis engine and provides a simple YAML interface for custom rules.
  \item \emph{Semantic (logic)}. \infer (v1.1.0)~\cite{facebook2022infer} and \codeql (v2.12.0)~\cite{github2021codeql} derive the computational model from the source code by interpreting the program's instructions.
  \infer is developed by Meta and uses separation logic to reason about heap-based pointer structures in C/C++ programs.
  \codeql is developed by GitHub and uses a deductive database to store program facts.
  Vulnerability patterns are encoded as declarative queries to the database.
  \item \emph{Semantic (symbolic execution)}. \clangsa (v12.0.0)~\cite{llvm2020clang} follows a symbolic execution-based approach. Code fragments are symbolically executed to construct a symbolic state. It checks whether there is a satisfiable path to a dangerous state where security checks are violated. \clangsa aims at undefined behavior and dangerous code constructs, but can be extended with other types of checkers~\cite{clang_analyzer_checker_developer_manual}.%
\end{itemize}
Except for \flawfinder and \semgrep, all analyzers need to be integrated into the program's build process. 
Apart from \codeql and \semgrep, all analyzers have a fixed set of rules that may be enabled or disabled. 
\codeql and \semgrep support the definition and addition of new detection rules.

\begin{sloppypar} %
\cparagraph{Fuzzing tools}
As the basis for our fuzzing tool set, we use all fuzzers available in \magma v1.1:
\afl~\cite{zalewski2017american}, \aflfast~\cite{bohme2016coveragebased}, \aflppthree~\cite{fioraldi2020afl}, \angora~\cite{chen2018angora}, \entropic~\cite{bohme2020boosting}, \fairfuzz~\cite{lemieux2018fairfuzz}, \honggfuzz~\cite{google2010honggfuzz}, \libfuzzer~\cite{serebryany2016continuous}, \moptafl~\cite{lyu2019mopt}, \parmesan~\cite{osterlund2020parmesan} and \symccafl~\cite{poeplau2020symbolic}.
We use the pre-defined settings for these fuzzers and no external sanitizers.
Additionally, we augment this set with a more recent version of \aflpp~\cite{fioraldi2020afl} (4.08c) and \libafl~\cite{crump2023libafl} (version 0.13.1), a libAFL-based replacement for the now-deprecated libFuzzer.
For \aflpp, we refer to the more recent version unless specified otherwise.
\end{sloppypar}

\section{Empirical Evaluation}
\label{sec:quant-eval}

We begin our evaluation by quantitatively comparing the selected \ac{sast} and fuzzing tools in terms of their bug finding capabilities, specifically focusing on true positives. 
More specifically, we address the following research questions:
\begin{description}[leftmargin=1em, nosep]
  \item[RQ.1 True Positives] How many bugs in the \magma and \cgc benchmarks are identified by each bug finding tool and approach?
  \item[RQ.2 Complementarity] Do static analysis and fuzzing complement each other, both within and across approaches, or do they essentially identify the same set of bugs?
  \item[RQ.3 Bug Types] Are static analysis tools more effective at finding certain types of bugs compared to fuzzers, and vice versa?
  \item[RQ.4 Overhead] How much manual effort does a tool impose on a developer to extract actionable results (e.g., related to false positives or deduplication of findings)?
\end{description}

\subsection{Experimental Design}
\label{sec:quantdesign}

We set up an evaluation platform to fairly compare all bug finders.
To prevent negatively influencing \ac{sast} tools, we remove \magma's canaries for static analysis, leaving only the vulnerable code.

\cparagraph{SAST configuration}
\Ac{sast} tools offer a variety of configuration options that enable or disable specific checks, allowing users to fine-tune the analyzer for particular objectives (e.g., finding vulnerabilities instead of code smells).
We configure each tool to optimize its performance and also enable all security-relevant options.
Since code quality measurements are outside the scope of this work, we disable code smell reporting where possible.
For verifiability, we include this configuration mapping in our artifact.

\cparagraph{Quantifying false reports}
Thanks to ground truth data, we can automatically evaluate the false \emph{negative} rate, i.e., the number of expected \acsp*{cve} \emph{not} found.
However, we \emph{cannot} reliably evaluate the false \emph{positive} rate, i.e., the number of invalid bug reports per tool.
Since false positives are an issue frequently associated with static analysis, we nevertheless approximate this overhead by studying the number of bug reports generated per every expected CVE (RQ4), and we manually classify all 424 \ac{sast}-generated bug reports in functions with known \acsp{cve} in \autoref{sec:qual-manual-validation}.

\cparagraph{Identifying whether a CVE is discovered}
To identify bugs discovered by fuzzing, we use the canaries provided by \magma and
manual crash inspection for \cgc, as it has no canaries.
For static analysis, we take a different approach since static analyzers may report unrelated bugs in functions containing ground-truth bugs. 
For example, a tool might report a null pointer dereference when we expect a buffer overwrite.
Following \etal{Lipp}~\cite{lipp2022empirical}, we identify detected bugs by comparing the reported vulnerability type (\acs{cwe}) with the expected type. 
We define expected \acsp{cwe} based on \magma's provided data and \cgc's vulnerability descriptions.
Because \acsp{cwe} are somewhat ambiguous, we also accept sufficiently similar \acsp{cwe} as matched.
To this end, we map each expected \acs{cwe} to up to 27 related \acsp{cwe}.
For example, if we expect an improper validation of array index (CWE-129), we accept reports of improper input validation (CWE-20) or out-of-bound reads (CWE-125).
A full list is available in the Appendix (\autoref{tbl:similar-cwes}).

\cparagraph{Setup and Infrastructure}
We rerun the \magma experiments on v1.1 instead of using data from the original paper~\cite{hazimeh2020magma} since v1.0 contained unreachable bugs that were removed in later versions~\cite{lipp2022empirical}.
Following recommendations from \etal{Klees}~\cite{klees2018evaluating} and \etal{Schloegel}~\cite{schloegel2024sok}, each fuzzer runs for \emph{48~hours} with \emph{10 trials}.
We increase the corpus size to \SI{100}{\gibi\byte} and restrict workers to physical cores to avoid SMT-related variations.
The experiments run on seven identical machines with 52-core Xeon Gold 6230R CPUs and \SI{188}{\gibi\byte} RAM.
All machines use Ubuntu 22.04 LTS with kernel version 5.15.
For static analysis, we run the experiments in an LXC container with a Xeon Gold 6248R CPU using 16 cores and \SI{128}{\gibi\byte} RAM.
The container uses Ubuntu 20.04 and kernel version 6.5.

\subsection{RQ.1 True Positives}

\begin{figure*}
  \subcaptionbox{Fuzzer-detected bugs in different trials\label{fig:fuzzer-detailed}}{%
    \includegraphics{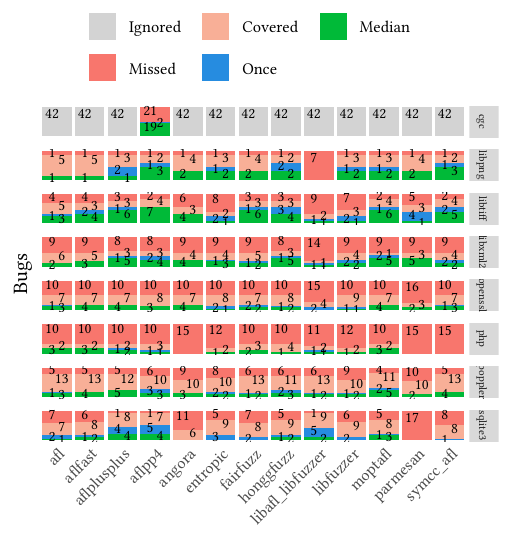}
    \vspace{-0.25cm}
  }
  \subcaptionbox{\ac{sast} reports\label{fig:sast-detailed}}{%
    \includegraphics{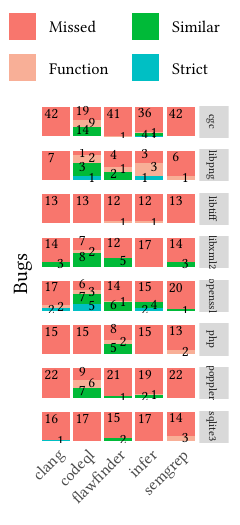}
    \vspace{0.25cm}
  }
  \caption{Flags per tool and target in \magma and CGC.} %
  \label{fig:fuzzer-sast-detailed}
\end{figure*}

We evaluate each fuzzer using ten 48-hour trials on the seven \magma programs. 
For \cgc, we only evaluate \aflpp with ten 48-hour trials due to the required manual verification of results.
\autoref{fig:fuzzer-detailed} shows the number of bugs that were \emph{Covered}, found at least \emph{Once}, found in at least six trials (\emph{Median}), or \emph{Missed} across all runs.
We use \emph{Once} for RQ2 and RQ3 to assess the tools' potential performance and \emph{Median} in RQ4 to evaluate expected real-world performance.%
The results for the five static analyzers are shown in \autoref{fig:sast-detailed}. 
As described in \autoref{sec:quantdesign}, we classify detection accuracy into multiple categories: a vulnerability is considered detected if its vulnerability type (CWE) exactly matches the expected CWE (\emph{strict}) or is sufficiently similar (\emph{similar}), based on our CWE similarity criteria (\autoref{tbl:similar-cwes}). 
For undetected bugs, we track whether the analyzer reported any bug in the vulnerable function (\emph{function}) or missed it entirely (\emph{missed}).

\cparagraph{Fuzzers}
Looking at \autoref{fig:fuzzer-sast-detailed}, among the 112 bugs in \magma, the fuzzer \aflpp finds 37 bugs while the static analyzer \codeql detects 31. 
This similar performance between the best tools in each category suggests that both fuzzing and static analysis have similar potential to detect bugs, though there still remains room for improvement for both techniques.
For the fuzzers (\autoref{fig:fuzzer-detailed}), the number of bugs covered and found varies greatly between projects, but only slightly between fuzzers.
In \libxml, \openssl, and \php, many undetected bugs are not even covered by the fuzzer. %
For \poppler, \sqlite, and \libpng, many bugs are covered but never found, suggesting specific failure conditions that fuzzers struggle to overcome.

\cparagraph{SAST Tools}
The static analyzers (\autoref{fig:sast-detailed}) show significant variation in their performance.
Their effectiveness differs based on their semantic reasoning capabilities.
\codeql found the most security bugs, likely due to its extensive reasoning capabilities.
On the \magma dataset, \flawfinder ranks second but with a notable performance gap.
\flawfinder detects potential vulnerabilities by searching for dangerous C standard library functions like \texttt{strlen}, \texttt{strcpy} or \texttt{sprintf}, without reasoning if the function arguments are properly validated.
This explains \flawfinder's poor performance on the CGC benchmark, which rarely uses standard library functions.
Across all analyzers, despite generating 44,332 bug reports, most functions containing security bugs receive no flags, even for unrelated bug types (see \emph{Function} in \autoref{fig:sast-detailed}).

\begin{figure*}[htb]
  \centering
  \includegraphics{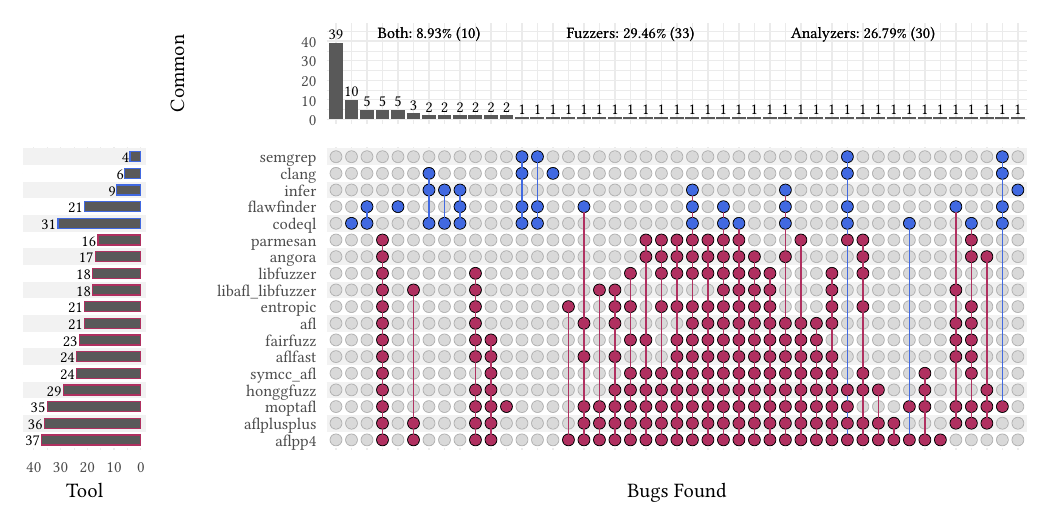}
  \vspace{-0.8cm}
  \caption{
    Intersection plot~\cite{lex2014upset} of \magma bugs found by each \textcolor{RoyalBlue}{analyzer} or \textcolor{Maroon}{fuzzer}. %
    Every row in the intersection matrix represents the set of bugs found by a tool, and every column is a set of bugs commonly detected by a group of tools.
    A color-filled circle indicates that a tool is part of a tool group.
    Bar plots visualize the set sizes of tools (left) and tool groups (top).
  }
  \label{fig:upset-overall}
\end{figure*}

\subsection{RQ.2 Complementarity}

\autoref{fig:upset-overall} shows how static analysis and fuzzing complement each other in an UpSet (or intersection) plot \cite{lex2014upset}.
Out of 112 \acsp*{cve} in \magma, 39 are not found by any approach, 33 are found only by fuzzers, 30 are found only by static analyzers, and 10 are found by both approaches. 
These results indicate that both approaches find different types of bugs, making them complementary.

\cparagraph{Fuzzers}
Fuzzers tend to find the same \acsp*{cve}. 50\% of the \acsp*{cve} are found by at least five fuzzers, and a quarter of those found by \emph{any} fuzzer are found by \emph{all} fuzzers.
This homogeneity suggests that using a single well-performing fuzzer like \aflpp, which finds 37/43 \acsp*{cve}, captures most vulnerabilities detected by fuzzing.

\cparagraph{SAST Tools}
Static analysis tools discover less similar sets of bugs.
\codeql discovers a large set of bugs exclusively, and also the most in total.
Only \flawfinder finds more than one \acs{cve} not already detected by \codeql.
Since \codeql significantly outperforms other tools, using \codeql alone will provide nearly all findings.

\subsection{RQ.3 True Positives Across Bug Types}

\begin{table}
  \centering
  \small
  \begin{threeparttable}
    \caption{Vulnerability Type Names and Their Description}
    \label{tbl:codenames}
    \begin{tabularx}{\linewidth}{llX}
      \toprule
      Name                            & CWE          & Description                                         \\
      \midrule
      \textsc{Bounds}                 & 119\tnote{*} & Improper Restrictions [of Buffer Operations]        \\
      \textsc{Resource}               & 664\tnote{*} & Improper Control of a Resource [\dots]              \\
      \textsc{Math}                   & 682          & Incorrect Calculation                               \\
      \textsc{Logic}                  & 691          & Insufficient Control Flow Management                \\
      \textsc{Except}                 & 703          & Improper [\dots] Handling of Exceptional Conditions \\
      \textsc{Data}                   & 707          & Improper Neutralization [of Data]                   \\
      \textsc{Style}\tnote{$\dagger$} & 710          & Improper Adherence to Coding Standards              \\
      \textsc{Other}\tnote{$\dagger$} & ---          & Remaining CWE pillars (284, 435, 693, 697)          \\
      \bottomrule
    \end{tabularx}
    \begin{tablenotes}
      \item[*] We split CWE 664 s.t. all out-of-bounds issues (subsumed under CWE 119) are separately counted, as the majority of bugs fall in this category.
      \item[$\dagger$] These vulnerability types are included for completeness, but are not central to our analysis.
    \end{tablenotes}
  \end{threeparttable}
\end{table}

\begin{figure}[htb]
  \centering
  \includegraphics{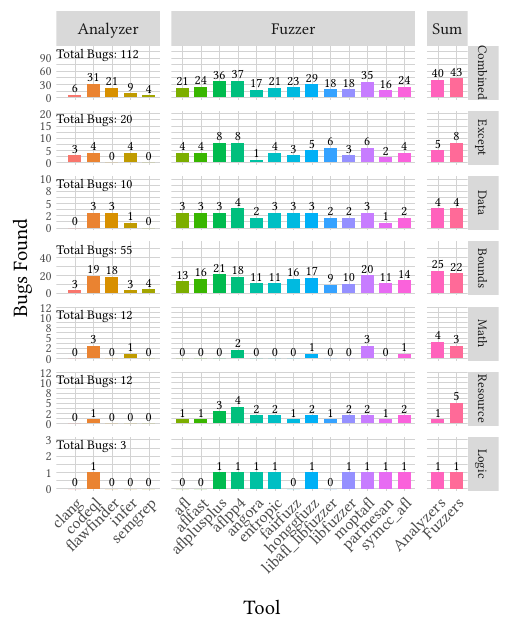}
  \vspace{-0.6cm}
  \caption{Detected \magma \acsp*{cve} per tool across bug types.
    For fuzzers, a bug is found if it is detected \emph{once}.
    Bug types (see \autoref{tbl:codenames}) not covered by \magma are omitted.
  }
  \label{fig:found-totals}
\end{figure}

\autoref{fig:found-totals} shows how many of \magma \acsp*{cve} are detected across six top-level vulnerability types (\acsp*{cwe}).
These vulnerability types are code names for the \emph{pillar} CWE IDs based on the \emph{Research Concepts} CWE view~\cite{mitre2026cwe}, as explained in \autoref{tbl:codenames}.
Note that we exclude \cgc here, since we only have results for one fuzzer.
For static analyzers, we consider a bug detected if its specific vulnerability type (\acs*{cwe}) is similar to the reported vulnerability type (see \autoref{sec:quantdesign})

The rows in \autoref{fig:found-totals} show results for each vulnerability type and their combined total. The columns show results per tool group and their sum total.

\cparagraph{SAST Tools}
Static analyzers perform best at detecting \emph{Bounds} vulnerabilities and show strong results for \emph{Data} vulnerabilities. %
They achieve moderate success with \emph{Except} and \emph{Math} vulnerabilities. 
\emph{Resource} vulnerabilities are challenging for static analyzers to detect.
This could be due to the complex global state typically required for e.g., temporal memory safety bugs to surface, which the intra-procedural reasoning of the \ac{sast} tools fails to capture.
Individual static analyzer tool performance varies significantly across analyzers. 
\codeql performs best overall, though other tools match its performance for specific bug types. 
\flawfinder ranks second in overall performance but fails to detect certain bug classes.
\clangsa, \infer, and \semgrep perform notably worse than the leading tools, primarily due to their limited detection of bounds-related vulnerabilities and complete misses of certain bug types.

\cparagraph{Fuzzers}
Fuzzers are most effective at detecting \emph{Except}, \emph{Data}, \emph{Bounds}, and \emph{Resource} vulnerabilities. 
They are less effective for \emph{Math} vulnerabilities, which might be due to the need to generate specific input values to trigger a bug.
We find that individual fuzzer performance is remarkably similar across different bug classes:
There is no fuzzer that excels at one particular bug type, but fails on another.
As they were all using the same sanitizer, fuzzers' bug detection capabilities depend on their input generation and selection strategies, which seem to generalize well across bug types.

\subsection{RQ.4 Overhead}
\label{sec:quant-overhead}

\begin{figure}
  \centering
  \includegraphics{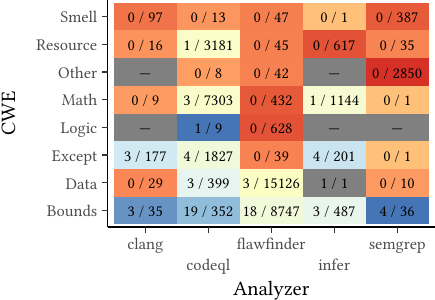}
  \caption{
    Number of discovered CVEs / generated \ac{sast} bug reports across vulnerability types in \magma.
    The color represents the ratio between reports and bugs.
    A blueish color indicates that true positives were found, while a reddish color indicates that none were found.%
  }
  \label{fig:sast-ratio-magma}
\end{figure}

After running automated bug finding tools, developers must investigate the results.
Since this investigation cannot be fully automated, tools should minimize developer overhead.
To quantify this overhead, for \ac{sast} tools we analyze the ratio of bug reports to expected bugs in the codebase and for fuzzers, which normally do not have false positives, we discuss other factors influencing overhead.

\cparagraph{SAST Tools}
For static analysis tools, the main overhead is manually filtering generated bug reports, as these tools are known to generate many false positives~\cite{beller2016analyzing, gosain2015static, imtiaz2019challenges}.
In~\autoref{fig:sast-ratio-magma}, we report the number of bug reports per expected \acs*{cve} across the eight high-level vulnerability types in \magma.
While we expect to find 112 \acsp*{cve}, the five analyzers generate 44 \emph{thousand} bug reports.
\clangsa generates fewer reports out-of-the-box, while other tools produce an impractical number of reports for developers to investigate (at least in our configuration).
Most importantly, \emph{only 0.24\%} of reports from \codeql, the best performing tool regarding true positives, correspond to expected ground truth bugs.
However, quantifying the exact number of false positives is difficult since we do not know how many bugs exist in \magma beyond the known forward-ported bugs.
Some of the 44k reports may indicate true positives, but analyzing all reports would be infeasible.
Managing large numbers of potentially invalid bug reports remains an active research area in software engineering and programming languages~\cite{christakis2016what,johnson2013why,imtiaz2019challenges,lipp2022empirical,sadowski2018lessons}.
Looking at \autoref{fig:sast-ratio-magma}, we can see that tools generate varying numbers of bug reports across vulnerability types.
For example, \codeql reports many \emph{Math} bugs across datasets.
Similarly, \flawfinder produces many reports for \emph{Data} and \emph{Bounds} bugs---these account for \emph{all} 21 bugs it finds in \magma. %
This is due to the syntactic nature of the tool: \flawfinder reports usage of a set of dangerous functions from the standard library. %
Thereby, it will miss bugs in non-standard functions and report even safe usages of dangerous functions.

\cparagraph{Fuzzers}
While fuzzers may miss bugs in individual runs due to their non-deterministic nature, they generally do not produce false positives (depending on the sanitizer).
However, when a fuzzer normally detects a crash, developers have to manually identify its root cause, a potentially time-intensive task.
Furthermore, fuzzers often generate multiple crashing inputs that are caused by the same underlying issue.
While these inputs are not false positives, deduplication remains challenging, though recent research has shown promising results in this area~\cite{jiang2021igor, kallingaljoshy2022fuzzeraid}.
Our evaluation relies on \magma's built-in triggers that simulate a perfect sanitizer, but, in contrast to real sanitizers, do not induce a crash.
Therefore, we cannot realistically quantify the number of duplicated crashes in our dataset.
Nevertheless, from the developer's perspective, this is no concern: Since all crashes represent true positives, they can arbitrarily pick and fix one of them, (programmatically) check which inputs still crash, and proceed to investigate this remaining subset.

\section{Qualitative Evaluation}
\label{sec:qual}

Beyond the quantitative evaluation, we aim to gain a deeper understanding of the utility of static analyzers and fuzzers.
We start by validating the results from our automated evaluation.
Afterward, we discuss two qualitative aspects using examples from our study:
First, we analyze the types of bugs found by both approaches, the complexity of reasoning required, and how bug detection depends on the execution environment.
Second, we evaluate the tools in terms of setup effort, configuration options, and resource requirements.
Finally, we look at integration points into development workflows and examine the real-world adoption of security tooling.

\subsection{Manual Validation}
\label{sec:qual-manual-validation}
To get a qualitative impression of the \ac{sast} tools' reports, we manually check only the static analyzer reports in functions changed by \magma's bug insertion patches.
This filters out reports that are not relevant with respect to the target bugs and reduces the total number of reports from 44,231 to 424,
which makes manual analysis feasible.
While this selection does not allow us to quantify the analyzers' false positive rate, it serves as a validation for our custom bug-report matching.
For each report, we are only interested in whether it describes the inserted bug close enough that a code reviewer could identify the bug.

For each \magma bug, we mark source code lines where tools reported bugs based on results from the five static analyzers in our empirical analysis (\emph{report}).
We also mark lines that caused the actual bug according to \magma (\emph{fault}) and how the bug was patched (\emph{patch}).
We generate a table containing 525 report-bug pairs with links into the annotated codebase for inspection.

\cparagraph{Manual Labeling}
Two researchers with expertise in automated bug finding and system building classified bug reports as true or false positives
with regard to known \magma bugs using the following task description:
``Open the code base at the given bug location. Investigate for each nearby \acs*{sast} flag
(i.e., flags in the same function) if it identifies the given bug.''
To mitigate threats to internal validity, we follow the standard coding protocol from grounded theory~\cite{glaser2017discovery}.
Each researcher independently assigned labels, spending approximately twenty hours.
The researchers then compared their results, finding they disagreed on 8\% of the labels.
All disagreements were resolved during multiple sessions taking another four hours in total.
We publish the meeting protocol to ensure verifiability.

\cparagraph{Results}
Out of 424 static analysis reports, only 35 (8.25\%) were manually confirmed to match a known CVE.
Out of 112 \magma bugs, only 12 were detected by any static analyzer in our study.
\codeql alone correctly detects 8 bugs, \flawfinder detects 5 bugs, \clangsa detects 3 bugs, while \infer and \semgrep each detect 2 bugs.
Consequently, while our automatic bucketing overestimates the tools' true positive rate by a factor of around three (3.3x),
we can see that it otherwise paints a realistic picture of the tools' capabilities.
We discuss all confirmed-positive findings in detail in \autoref{sec:appendix-manual}.

\subsection{Limits to Bug Discovery}
\label{sec:qual-limits}

Static analyzers and fuzzers detect different types of bugs due to their distinct approaches: fuzzers \emph{execute} programs, while static analyzers \emph{reason} about them.
In the following, we discuss the conceptual limits of each technology.

\cparagraph{Types of bugs}
Static analysis and fuzzing detect different types of bugs.
Bugs found by \emph{fuzzing} centrally depend on bug oracles (i.e., sanitizers).
While there are good sanitizers for memory safety~\cite{serebryany2012addresssanitizer, stepanov2015memorysanitizer, theclangteam2013undefinedbehaviorsanitizer}, certain bugs are difficult to detect---particularly logic bugs, which require inferring expected program behavior.

\emph{Static analysis} uses bug oracles encoded as patterns that are either hard-coded and toggleable (as in \flawfinder, \clangsa, and \infer) or fully customizable via rule definition languages (as in \codeql and \semgrep).
Based on these patterns, static analyzers can detect non-crashing semantic bugs affecting program correctness, such as SQL injections, information leaks, and logic bugs---issues that fuzzing cannot detect without a specialized oracle.

\cparagraph{Reasoning}
While static analysis predicts problematic states without execution, fuzzing can only detect issues in observed executions.
Consequently, fuzzers cannot find bugs in uncovered code.
Thus, carefully designed harnesses, good code coverage, and a meaningful seed corpus are crucial for fuzzing~\cite{klees2018evaluating}.
If integrity checks like checksums are hard to satisfy with random mutations, they need to be patched out for fuzzing to reach deeper code regions.

Static analyzers do not need to execute the program but instead are limited by their reasoning approach.
\emph{Syntactic analyzers} check syntactic patterns by comparing code fragments.
For instance, \flawfinder matches hard-coded regular expressions to flag problematic function calls (e.g., \texttt{sprintf}).
This approach is brittle since different syntactic representations can mask issues.
\emph{Semantic analyzers} can theoretically reason about program behavior without limitations.
However, they must handle large, multi-language codebases with complex invariants. %
We see this in our dataset: \openssl (626K LoC) is mainly C code, but 30\% consists of Perl preprocessing scripts, making analysis challenging for semantic analyzers.

\cparagraph{Environment}
Some bugs only manifest in a \emph{specific} program environment.
The program environment is determined by factors like the machine architecture, system calls, compiler flags, preprocessor directives, and program configuration.
\emph{Fuzzing} only finds bugs in the active environment.
For instance, in our setup, all programs are compiled for x86\_64, preventing fuzzers from detecting bugs that only occur on x86\_32.
For example, \magma bug MAE004 / PHP002 (\autoref{lst:php002} in \autoref{sec:appendix-data} shows the relevant code) is caused by overflowing a 32-bit user-controlled integer on a 32-bit system.
On 64-bit systems, however, the value is upcast to 64 bits, preventing the overflow and hiding the bug.
Thus, none of the fuzzers triggered this bug.
\emph{Static analysis} can, in theory, reason about programs under all possible environments and configurations.
Even for functions with unknown implementations, static analysis can model arbitrary return values.
However, this generality leads to more false positives.
To increase precision, semantic analyzers typically integrate into the build process to capture more information.
In turn, this makes them susceptible to the same environmental limitations as fuzzers.  %

\subsection{Usability and Resources}
\label{sec:qual-usability}

The adoption of bug finding tools depends not only on their effectiveness, but also on their usability.
We discussed the challenges of bug deduplication for fuzzing and false positive reports for static analysis in \autoref{sec:quant-overhead}.
In this section, we examine three additional qualitative usability factors: (i) setup complexity, (ii) configurability, and (iii) resource requirements.

\cparagraph{Setup}
Before using an automatic bug finding tool, it must be set up for the target program.
\emph{Static analysis} requires minimal target-specific setup.
Syntactic analysis tools like \flawfinder or \semgrep have no up-front cost and can directly analyze source code.
However, semantic analysis tools like \clangsa, \infer, and \codeql need integration into the project's build process. %
In our experience, this integration is straightforward, requiring only the build command to be passed to the static analyzers.

\emph{Fuzzing} has higher upfront costs.
The fuzzer's compiler must be integrated into the build process to add coverage instrumentation and possibly sanitizer instrumentation. %
Additionally, fuzzing requires a fuzz harness, which is often time-consuming and requires project-specific knowledge.
Reducing this effort through automatic harness generation (e.g., using recent advances in code generation models) is a topic of ongoing research~\cite{zhang2024how,yu2024promptfuzz}.

\cparagraph{Tuning}
Developers may want to fine-tune bug finding tools for their specific projects. While tools usually aim to be general-purpose, projects often have differing requirements.
On the one hand, \emph{static analysis} tools are usually highly configurable to support tuning for specific codebases.
This configurability helps manage their high false positive rate by allowing developers to disable certain analyses or ignore parts of the codebase. %
Additionally, tools that decouple their rule set from the analysis engine allow developers to encode custom bug patterns, enabling analysis specific to the codebase.
Because of the sheer size of this configuration space, we could not hope to meaningfully cover it in our experiments, and opted to over-approximate instead by enabling all available security-relevant analyses at the price of more false positives.

\emph{Fuzzers}, on the other hand, can be tuned further through dictionaries~\cite{fioraldi2020afl}, grammars~\cite{pham2020smart}, protocol awareness~\cite{pham2020aflnet}, improved seed corpora, and better fuzz harnesses~\cite{serebryany2016continuous}.
In practice, bug oracles are also a crucial aspect.
Besides generic sanitizers, these can be made target-specific e.g. via differential testing.

\cparagraph{Hardware Resources}
Hardware resources are an important consideration when adopting automatic bug finding tools.
Fuzzers are known to require lots of hardware resources, indeed, a typical fuzzer could be run forever since there is no inherent stopping rule.
In contrast, \emph{static analysis} terminates once analysis results are generated.
To compare these differences concretely, we calculate the approximate cost per bug found for each tool on the \magma bugs.
Note that this is a coarse approximation given that many factors need to be taken into account, and as our results show, the cost can heavily vary depending on the analyzed subject.

Assuming energy costs of \SI{40}{\cent\per{\kilo\watt\hour}}, we calculate the cost per core hour.
The fuzzing CPUs (26 cores/socket, TDP \SI{150}{\watt}~\cite{intel_xeon_6230r}) cost \SI{0.23}{\cent\per\hour\per\text{core}}.
The static analysis CPUs (24 cores/socket, TDP \SI{205}{\watt}~\cite{intel_xeon_6248r}) cost \SI{0.35}{\cent\per\hour\per\text{core}}.
We estimate the cost per bug by multiplying these rates with the tool runtime and number of harnesses for fuzzing, then dividing by bugs found.
We recorded runtimes for each static analyzer and subject (see \autoref{tbl:sast-runtimes} in \autoref{sec:appendix-data}).
For this calculation, we count analyzer flags using the \emph{similar} matching described in \autoref{sec:quantdesign} for static analysis and the \emph{median} run for fuzzers.
This results in the following costs: \flawfinder is cheapest at \SI{0.0003}{\cent\per\text{bug}}, followed by \semgrep at \SI{0.005}{\cent\per\text{bug}}.
Semantic analysis tools require more power: \clangsa costs \SI{0.09}{\cent\per\text{bug}}, while \codeql (\SI{0.4}{\cent\per\text{bug}}) and \infer (\SI{1.9}{\cent\per\text{bug}}) are the most expensive \ac{sast} tools.
Fuzzing is even more expensive, with the best performer (\aflpp) costing \SI{58.5}{\cent\per\text{bug}}.%
\footnote{
  We are aware that these calculations depend on our evaluation parameters and targets and may not generalize perfectly.
  Practitioners might terminate \ac{sast} tools early when they appear stuck rather than waiting 24 hours.
  They may also run fewer parallel fuzzing campaigns or extend campaigns beyond 48 hours.
  Nevertheless, these numbers illustrate the relative resource usage differences between tools.
}
Note that fuzzing costs are expected to grow exponentially over time for a linear increase in found bugs~\cite{bohme2020fuzzing}.
All in all, we see that tools that detect more vulnerabilities in total also incur higher computational costs per bug.

\subsection{Bug Finding in the Development Process}
\label{sec:qual-pipeline}

Bugs are cheaper to fix when caught early in the development process, so OWASP recommends applying security tooling early (``shift left'' paradigm)~\cite{owasp2021application}.
In the following, we examine where fuzzers and \ac{sast} tools fit in the development pipeline and investigate how they are used in practice.

Due to its high runtime cost (discussed in the previous section), fuzz testing is typically deployed late in development on already \emph{released} software.
For example, OSS-Fuzz~\cite{serebryany2017ossfuzz} mostly tests the public main branches of popular open-source software~\cite{githubOssfuzz}.
Fortunately, our study suggests significant overlap between different fuzzers, making it less important to use multiple fuzzers.

In contrast, static analysis tools can be used earlier in the development pipeline and can easily be integrated into \emph{continuous integration} on popular code sharing platforms~\cite{hutchings2020code}.
As discussed in \autoref{sec:qual-usability}, tools can be fine-tuned or limited to recently changed code to reduce reports while still improving code quality incrementally.
However, our results presented in \autoref{sec:quant-overhead} suggests that \ac{sast} tools are not utilized in \magma projects, given the large number of reports generated.

Some bugs can be caught even earlier in the pipeline, \emph{at compile time}.
Using a current language version and enabling compiler warnings
can help to detect and prevent issues early---for example, the Linux kernel team upgraded from C89 to C11 in 2022 to support block-level variable declarations after a vulnerability~\cite{corbet2022moving}.
Addressing compiler warnings also reduces noise from subsequent static analysis.

\cparagraph{CI Pipeline Analysis}
Motivated by the insight that (at time of inclusion into \magma), some of the projects had apparently not used bug finding tools as part of the development process, we want to quantify the in-practice usage of such tools.
Notably, \etal{Beller}~\cite{beller2016analyzing} found in 2016 that almost half of the open source projects they surveyed used static analysis, but only sporadically.
To get a sense of how commonly fuzzers and \ac{sast} tools are used in the most critical 100 OSS projects today~\cite{wgsecuringcriticalprojects2023securing}, we manually analyzed all \data[ci-stats-data][total] OSS projects that are hosted on \github. %
There, we automatically search for tool names in the repository's CI configuration, manually double-checking the results.
A project also counts as fuzzed if it has a harness in OSS-Fuzz.
We found that
\data[ci-stats-data][both] of the \data[ci-stats-data][total] analyzed projects use both fuzzing and \ac{sast} tools,
\data[ci-stats-data][only-fuzzer] use only fuzzing,
\data[ci-stats-data][only-sast] use only \ac{sast} tools, and
\data[ci-stats-data][neither] projects use neither of the two methods.

\section{Discussion}
\label{sec:discussion}

Based on our qualitative and quantitative analyses, we provide advice for software project maintainers looking to improve security, and discuss potential improvements for fuzzing and static analysis research.

\subsection{Recommendations for Project Maintainers}
Since \ac{sast} tools and fuzzers find different types of bugs with minimal overlap, they complement each other.
While fuzzers are generally recommended if computational resources are available, combining them with \ac{sast} tools can be beneficial despite the additional manual effort required.
Given the relatively low computational expenses that we see (in \autoref{sec:qual-pipeline}) for running \ac{sast} tools, they should be run early in the pipeline (e.g., on every commit).

Specifically, we would recommend semantic over syntactic \ac{sast} tools due to the enormous noise introduced by syntactic tools flagging all occurrences of certain code patterns.
Even though a single \ac{sast} tool (\codeql) provides nearly all true positives among the tested tools, without any mitigation, the number of bug reports would make its use impractical on existing codebases.
One mitigation is to analyze only newly added or modified code to avoid being overwhelmed.
Additionally, most tools allow for customization which can be used to disable or adapt detection rules.
For example, as the largest number of reports for \codeql stem from checks for \emph{Math} bugs, disabling these reports would more than halve the number of total reports while only minimally affecting true positives.

In contrast, fuzzers find more bugs in our evaluation and only produce true positives, but need more computational resources and require a more elaborate initial setup, such as writing harnesses and build integration.
It is also possible to use fuzzing in continuous integration~\cite{clusterfuzzlite2025}.
Fortunately, a single tool (\aflpp) will find nearly all bugs that fuzzers would find.

Taking a broader perspective, the fact that roughly 50\% of the \magma bugs are not detected by \emph{any} tool%
\footnote{According to our manual validation in \autoref{sec:qual-manual-validation}, \ac{sast} tools only detected 12 bugs.} hints at a fundamental limitation of memory-unsafe languages.
C and C++ rely heavily on the programmer's diligence to prevent memory vulnerabilities, as ownership cannot be checked at compile time for these languages~\cite{borrowingTrouble2021}.
In contrast, modern systems programming languages address this issue directly through their design: languages like Go use automatic memory management, while Rust enforces explicit ownership checking at compile time.
Although these approaches involve trade-offs, they effectively eliminate memory safety violations through \emph{language design} rather than relying on external tools to catch such errors after the fact.

\subsection{Future Work}
\label{sec:lessons-learned}

Comparing \ac{sast} tools and fuzzers, the found bugs are quite disjunct.
While not entirely surprising, this is an interesting finding, that should motivate future work to combine the two approaches.
For example, there already is work on reducing false positive reports of \ac{sast} tools by using fuzzing to double-check the results~\cite{murali2024fuzzslice}.
It might also be feasible to use the extensive number of rules available in \ac{sast} tools to guide fuzzing (explored in~\cite{lipp2024sast}). %

Fundamentally, as we see in \autoref{sec:qual-manual-validation}, some bugs found by \ac{sast} tools are not even findable by fuzzers.
This hints to the open problems of defining better bug oracles (sanitizers) and harnesses.
Regarding \ac{sast} tools, in line with previous research, we see a vast amount of false positives, which will pose the main hurdle to adoption of \ac{sast} tools in practice.

\cparagraph{Benchmarks for Static Analysis}
We find that a fundamental problem for static analysis research is the lack of real-world benchmarks. 
The fuzzing community has Magma~\cite{hazimeh2020magma} and Fuzzbench~\cite{metzman2021fuzzbench}, which provide benchmarks based on real-world programs.
While not perfect, as manual bug selection can introduce bias, these benchmarks contain realistic targets.
However, to our knowledge, no equivalent exists for \ac{sast} tools, which are typically evaluated on small artificial code snippets like Juliet~\cite{nist_2017_juliet}.
Such a benchmark does not represent realistic programs, which can have deeply nested code, as well as complex and stateful bugs.

Creating a real-world benchmark with \ac{sast} tools in mind poses a number of challenges.
Firstly, the analyzers' outputs differ, even if using the standardized \ac{sarif} format.
We worked around this problem by creating manual mappings from sanitizer outputs to \acsp{cwe}.
However, \acsp{cwe} have limitations: it is often unclear which \acs{cwe} to assign to a bug, since \acsp{cwe} overlap and can be very generic.
While \acsp{cwe} are hierarchically organized, the resulting clusters are too coarse, which required us to develop a stricter relatedness clustering for our study (see \autoref{sec:appendix-data}, \autoref{tbl:similar-cwes}).

Secondly, tools report line numbers to mark bug locations, which we mapped to functions for coarse-grained analysis. 
Information on a level of affected variables would provide better precision. 
Complex bugs may appear in multiple code locations, potentially far from their root cause, and a fair benchmark should handle these cases. 

Lastly, a good benchmark design should also prevent gaming the system---tools that flags every line or variable provide no value.
At the same time, it should not punish tools for reporting bugs in the dataset that were unknown at the point of its compilation.
In our evaluation, we had to accept this source of over-approximation, erring on the side of static analyzers.

\section{Conclusion}
\label{sec:conclusion}

In this work, we present the first systematic comparison of static analysis and fuzzing for addressing typical C/C++ security weaknesses in real-world scenarios.
Quantitatively, we find that fuzzers detect slightly more vulnerabilities, with relatively little variation between tools.
This consistency is expected, as most modern fuzzers are derivatives of \afl.
Notably, fuzzers with sophisticated techniques, such as \angora or \symccafl, often perform worse than conceptually simpler approaches.
In contrast, static analyzers show greater performance variation.
Notably, \codeql detects more true positives than other tools, likely due to its extensive library of security analysis queries and the resources behind its commercial development.
An unexpected finding is that fuzzing and \ac{sast} tools find mostly disjoint sets of bugs.
This suggests that, to maximize true positive results, developers should use both types of tools in tandem for a more comprehensive bug detection strategy.

While \codeql leads the \ac{sast} field in true positive reports, these come at the cost of many false positives.
Indeed, all static analysis tools suffer from a very high false positive rate, well above the 15--20\% threshold considered acceptable by developers~\cite{christakis2016what}.
While customizing static analysis tools for specific projects can improve precision (e.g., by disabling Math and Resource checks in \codeql), our data suggests the resulting rates would still exceed this threshold.
Note that we focus on C/C++ code in this work, which is particularly challenging for static analysis.
In contrast, fuzzers demonstrate more practical utility despite requiring greater computational resources and initial effort for writing fuzzing harnesses.
This can be attributed to their dynamic nature---unlike static analyzers, they do not reason about a program but only observe execution side effects, namely crashes, which prevents false positive reports.
Given that developers prefer investing more analysis time for more precise results~\cite{christakis2016what}, fuzzing seems to be a more suitable approach in practice.

We also investigated the adoption of \ac{sast} tools and fuzzers in security-critical open-source projects and found that most projects already use at least one of these approaches.
Given that no tool in our experiment detects more than 40\% of the bugs, it seems clear that the definitive solution for memory safety bugs lies in the transition to safer programming languages that prevent these dangerous vulnerabilities by construction.

\begin{acks}
We would like to thank Thorsten Holz and Marcel Böhme for their valuable advice, feedback, and discussions on this work.
Furthermore, we thank Nico Schiller, Nils Bars, Mahnur Asif and all our colleagues for their feedback on the draft.
We would also like to thank Joschua Schilling for his advice and Julian Rederlechner for his support with the manual analysis of analyzer reports.
\end{acks}

\newpage
\printbibliography

\appendix
\section{Additional Data and Discussion}
\label{sec:appendix-data}

In this section, we provide additional data from our experiments that complement the quantitative and qualitative evaluation sections.
We present the total number of \ac{sast} reports in \autoref{tbl:sast-report-count} and the runtime of the \ac{sast} tools used for the calculation of cost-per-bug in \autoref{tbl:sast-runtimes}.

As we describe in \autoref{sec:quantdesign}, we were unsatisfied with the granularity of the \ac{cwe} hierarchy for clustering bugs, hence we manually derived a \ac{cwe} bucketing for automatic evaluation of \ac{sast} reports.
This mapping can be found in \autoref{tbl:similar-cwes}.

\autoref{sec:qual-limits} describes a bug in PHP that can only surfaces on 32-bit systems.
The corresponding source code can be found in \autoref{lst:php002}. 

\subsection{Manually Confirmed True Positive Bugs}
\label{sec:appendix-manual}

As we describe in \autoref{sec:qual-manual-validation}, we manually validated all \ac{sast} bug reports in functions that contain a \acs{cve} in Magma.
We list all confirmed true positive bugs in \autoref{tbl:cve-review-result}.
In the following, we discuss our findings beyond the raw number of true positives.

\cparagraph{Bugs missed by bucketing}
Two of the true positive reports are missed by our \emph{Similar} bucketing strategy.
Bug AAH046 (\href{https://nvd.nist.gov/vuln/detail/CVE-2019-10873}{CVE-2019-10873}) is tagged as a \texttt{NULL} pointer dereference (CWE-486) issue in the CVE database.
This is wrong; in fact, this bug reads out-of-bounds (and in consequence, might trigger a segmentation fault due to an invalid dereference of the resulting pointer).
If it had the correct CWE assigned, our bucketing strategy would have counted the flag as matching the bug.
The second bug, MAE100 (\href{https://nvd.nist.gov/vuln/detail/CVE-2019-10873}{CVE-2019-10873}), was initially tagged as ``numeric error'' (CWE-189), which is a CWE category that is not meant to be assigned to bugs.
In the meantime, this has been corrected to ``integer overflow'' (CWE-190).
However, \flawfinder reports a buffer overflow on the \texttt{memcpy} that uses the overflown number (as it does with \emph{every} \texttt{memcpy}, irrespective of arguments).
Manually, we can classify this specific flag as a true positive, but our automatic \emph{Similar} bucketing deliberately does not map between integer overflows and buffer overreads, as they are semantically different.

\cparagraph{Bugs not detected by fuzzers}
Looking at the manually validated \ac{sast} flags in \autoref{tbl:cve-review-result}, we see that only two of the twelve identified bugs were detected by \emph{any} fuzzer in the quantitative analysis.
This further confirms our insight that the two approaches find an almost-disjunct set of bugs.
Six of the twelve bugs were at least covered by fuzzers, leaving four bugs covered, but not detected.
This is especially interesting as we can rule out sanitizers as the cause: \magma provides a perfect oracle; if the bug condition is fulfilled, it will be detected.
Consequently, we hypothesize that the bug condition in these four cases must be difficult to trigger.
The remaining six bugs were not even covered, showing the need for improved harnesses to increase code coverage.

\begin{table}[t]
  \centering
  \caption{Number of Reports Generated by SAST Tools}
  \label{tbl:sast-report-count}
  \begin{tabular}{lrrr}
    \toprule
    Analyzer    & \magma & \cgc & Total \\
    \midrule
    \clangsa    & 363    & 0    & 363   \\
    \codeql     & 13092  & 2208 & 15300 \\
    \flawfinder & 25106  & 1019 & 26125 \\
    \infer      & 2451   & 331  & 2782  \\
    \semgrep    & 3320   & 9    & 3329  \\
    \addlinespace
    Total       & 44332  & 3567 & 47899 \\
    \bottomrule
  \end{tabular}
\end{table}

\begin{table}[t]
  \centering
  \small
  \begin{threeparttable}
    \caption{CPU Core Time [hh:mm:ss] for SAST Tools}
    \label{tbl:sast-runtimes}
    \begin{tabularx}{\linewidth}{lr*{5}{>{\raggedleft\arraybackslash}X}}
      \toprule
      Subject & LoC  & semgrep & clang-scan     & codeql            & infer             & flaw\-finder \\
      \midrule
      libtiff & 80K  & 09      & 04:07          & 09:46             & 09:53             & 02           \\
      libxml2 & 201K & 22      & 14:37          & 01:20:54          & 42:28:18\tnote{1} & 05           \\
      libpng  & 63K  & 06      & 01:42          & 06:33             & 09:31             & 02           \\
      php     & 865K & 01:02   & 09:00          & 30:22:22\tnote{2} & 04:56:37\tnote{1} & 36           \\
      sqlite3 & 445K & 01:03   & 16:43          & 24                & 28                & 12           \\
      poppler & 290K & 02      & 01:27\tnote{3} & 31:29             & 20:21             & 06           \\
      openssl & 626K & 35      & 26:49          & 05:27:30          & 58:16             & 07           \\
      \midrule
      Total   &      & 03:19   & 01:14:25       & 37:58:58          & 49:03:24          & 01:10        \\
      \bottomrule
    \end{tabularx}
    \begin{tablenotes}
      \item[1] killed out-of-memory
      \item[2] timeout after 24 hours real time
      \item[3] failed to capture build system
    \end{tablenotes}
  \end{threeparttable}
\end{table}

\begin{table}[t]
  \caption{Custom \acs*{cwe} Bucketing Used in Our Analysis}
  \label{tbl:similar-cwes}
  \small
  \begin{tabularx}{\linewidth}{lX}
    \toprule
    Expected & Similar Vulnerability Types (\acsp*{cwe}) \\
    \midrule
    CWE-20  & 20, 119, 120, 125, 126, 129, 134, 170, 190, 196, 466, 786, 787, 788, 805, 823 \\
CWE-129 & 20, 119, 120, 125, 126, 129, 786, 787, 788, 805, 823                          \\
CWE-119 & 119, 120, 125, 126, 190, 196, 676, 788, 805                                   \\
CWE-120 & 119, 120, 125, 786, 787, 788, 805, 823                                        \\
CWE-121 & 119, 120, 121, 125, 190, 196, 468, 682, 786, 788, 823                         \\
CWE-125 & 119, 120, 125, 126, 190, 196, 468, 786, 787, 788, 805, 823                    \\
CWE-126 & 119, 120, 125, 126, 131, 190, 196, 786, 787, 788, 805, 823                    \\
CWE-787 & 119, 120, 125, 190, 196, 676, 786, 787, 788, 805, 823                         \\
CWE-131 & 119, 120, 125, 126, 131, 467, 468, 682, 786, 787, 788, 805, 823               \\
CWE-369 & 369, 457, 682                                                                 \\
CWE-190 & 20, 190, 196, 682                                                             \\
CWE-401 & 401, 404, 772, 775                                                            \\
CWE-415 & 415, 416, 672                                                                 \\
CWE-416 & 415, 416, 672                                                                 \\
CWE-457 & 457                                                                           \\
CWE-681 & 195, 196, 197, 681, 704, 1077                                                 \\
CWE-770 & 20, 119, 120, 125, 129, 134, 170, 190, 466, 770, 786, 787, 788, 789, 805, 823 \\
CWE-772 & 401, 404, 772, 775                                                            \\
CWE-476 & 457, 476, 690                                                                 \\
CWE-835 & 196, 682, 834, 835                                                            \\
CWE-670 & 617, 670                                                                      \\
CWE-755 & 703, 755                                                                      \\
CWE-754 & 252, 703, 754                                                                 \\
CWE-399 & 399, 400, 402, 770, 834, 835                                                  \\
CWE-664 & 399, 400, 402, 664, 770, 834, 835                                             \\
CWE-189 & 128, 190, 191, 193, 196, 369, 839, 1335, 1339, 1389                           \\
CWE-201 & 201                                                                           \\
CWE-824 & 457, 824                                                                      \\
CWE-822 & 119, 822                                                                      \\
CWE-611 & 611                                                                          
                   \\
    \bottomrule
  \end{tabularx}
\end{table}

\begin{table*}
  \begin{threeparttable}
    \caption{\magma Bugs with Confirmed True Positive \ac{sast} Reports and Their Fuzz-Detectability}
    \label{tbl:cve-review-result}
    \centering
    \small
    \begin{tabular}{llrlrlrcc}
      \toprule
      \multicolumn{3}{c}{Bug} & \multicolumn{4}{c}{Static Analyzer Flag} & \multicolumn{2}{c}{Fuzzers\tnote{$\dagger$}}      \\
      \cmidrule(lr){1-3} \cmidrule(lr){4-7} \cmidrule(lr){8-9}
      \magma ID & Subject  & CWE & File                         & Line\tnote{$*$} & Tool      & CWE & Cov.   & Det.   \\
      \midrule
      AAH002    & \libpng  & 416 & png.c                        & 4554          & \codeql     & 416 & \xmark & \xmark \\
      AAH024    & \libxml  & 119 & valid.c                      & 1326          & \semgrep    & 676 & \cmark & \cmark \\
                &          &     &                              &               & \codeql     & 120 &        &        \\
                &          &     &                              &               & \clangsa    & 676 &        &        \\
                &          &     &                              &               & \flawfinder & 120 &        &        \\
      AAH029    & \libxml  & 119 & valid.c                      & 1371          & \semgrep    & 676 & \cmark & \cmark \\
                &          &     &                              &               & \clangsa    & 676 &        &        \\
                &          &     &                              &               & \flawfinder & 120 &        &        \\
                &          &     &                              & 1376          & \semgrep    & 676 &        &        \\
                &          &     &                              &               & \clangsa    & 676 &        &        \\
                &          &     &                              &               & \flawfinder & 120 &        &        \\
                &          &     &                              & 1379          & \semgrep    & 676 &        &        \\
                &          &     &                              &               & \clangsa    & 676 &        &        \\
                &          &     &                              &               & \flawfinder & 120 &        &        \\
                &          &     &                              & 1382          & \semgrep    & 676 &        &        \\
                &          &     &                              &               & \clangsa    & 676 &        &        \\
                &          &     &                              &               & \flawfinder & 120 &        &        \\
      AAH034    & \libxml  & 119 & xmlstring.c                  & 463           & \codeql     & 196 & \cmark & \xmark \\
                &          &     &                              & 468           & \flawfinder & 120 &        &        \\
                &          &     &                              & 499           & \codeql     & 196 &        &        \\
                &          &     &                              & 504           & \codeql     & 196 &        &        \\
                &          &     &                              &               & \flawfinder & 120 &        &        \\
                &          &     &                              & 505           & \codeql     & 196 &        &        \\
                &          &     &                              &               & \flawfinder & 120 &        &        \\
      AAH046    & \poppler & 476 & splash/SplashXPathScanner.cc & 444           & \codeql     & 196 & \cmark & \xmark \\
      AAH054    & \openssl & 119 & crypto/bio/b\_print.c        & 307           & \codeql     & 196 & \cmark & \xmark \\
                &          &     &                              & 846           & \codeql     & 190 &        &        \\
      JCH221    & \sqlite  & 476 & shell.c                      & 13666         & \clangsa    & 476 & \xmark & \xmark \\
      MAE100    & \openssl & 189 & crypto/evp/encode.c          & 173           & \flawfinder & 120 & \xmark & \xmark \\
      MAE102    & \openssl & 787 & crypto/mdc2/mdc2dgst.c       & 53            & \flawfinder & 120 & \xmark & \xmark \\
                &          &     &                              & 54            & \codeql     & 196 &        &        \\
      MAE103    & \openssl & 476 & ssl/statem/statem\_clnt.c    & 3140          & \infer      & 476 & \cmark & \xmark \\
                &          &     &                              & 3140          & \codeql     & 476 &        &        \\
      MAE106    & \openssl & 476 & crypto/pkcs7/pk7\_doit.c     & 610           & \codeql     & 476 & \xmark & \xmark \\
      MAE112    & \openssl & 476 & crypto/x509/x509\_vfy.c      & 1095          & \infer      & 476 & \xmark & \xmark \\
      \bottomrule
    \end{tabular}
    \begin{tablenotes}
      \item[$*$] We use the line numbers from our stripped benchmark set with \magma's preprocessor macros removed.
      \item[$\dagger$] Is this bug covered / detected by \emph{any} fuzzer in our quantitative evaluation?
    \end{tablenotes}
  \end{threeparttable}
\end{table*}

\begin{lstlisting}[float,label=lst:php002,caption={Part of PHP's EXIF handling code that contains CVE-2019-9641~\cite{desilva2019sec}, which only surfaces if \texttt{size\_t} is 32 bits.}]
static bool exif_process_IFD_in_TIFF_impl(image_info_type *ImageInfo, size_t dir_offset, int section_index)
{   /* ... */
    // integer overflow if dir_offset=0xFFFFFFFF
    if (ImageInfo->FileSize >= dir_offset+2) { 
		int sn = exif_file_sections_add(ImageInfo, M_PSEUDO, 2, NULL);

        // read fails (unchecked) because dir_offset is larger than file size
		php_stream_seek(ImageInfo->infile, dir_offset, SEEK_SET);
		php_stream_read(ImageInfo->infile, (char*)ImageInfo->file.list[sn].data, 2);

        // ImageInfo->file.list[sn].data is uninitialized (CWE-908)
		int num_entries = php_ifd_get16u(ImageInfo->file.list[sn].data, ImageInfo->motorola_intel);
        /* ... */
    } else {
        exif_error_docref(/* ... */);
    }
}
\end{lstlisting}
\clearpage

\section{Additional Figures}
\label{sec:appendix-figures}

\begin{figure}
  \centering
  \includegraphics{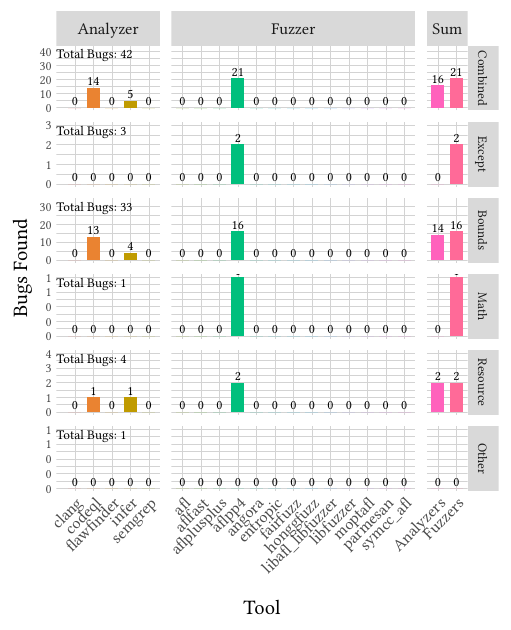}
  \vspace{-0.7cm}
  \caption{Total bugs found only for \cgc.}
  \label{fig:found-totals-cgc}
\end{figure}

\begin{figure}
  \centering
  \includegraphics{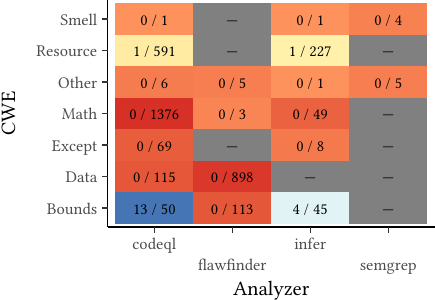}
  \caption{
    Number of discovered CVEs / generated \ac{sast} bug reports across vulnerability types for \cgc.
    Note that \cgc contains no \emph{Logic} bugs and \clangsa cannot be used on \cgc's custom standard library,
    hence we omit the corresponding column and row.
  }
  \label{fig:sast-ratio-cgc}
\end{figure}

\begin{figure}
  \centering
  \includegraphics{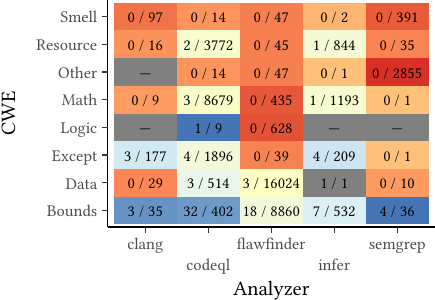}
  \caption{Number of discovered CVEs / generated \ac{sast} bug reports across vulnerability types for both \magma and \cgc.}
  \label{fig:sast-ratio-both}
\end{figure}

\begin{figure*}
  \centering
  \includegraphics{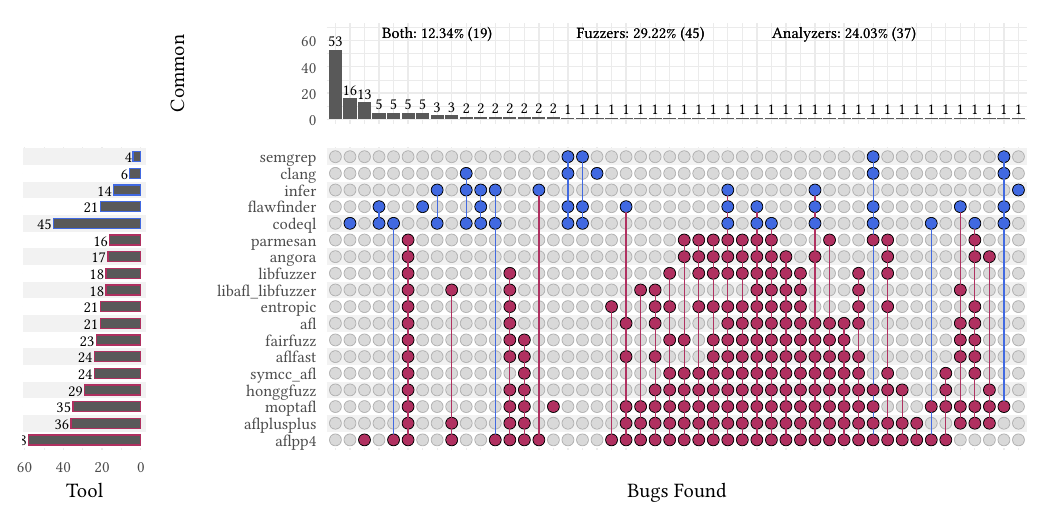}
  \vspace{-0.8cm}
  \caption{Upset plot to show overlap of bugs for both \magma and \cgc.}
  \label{fig:upsetr-overall}
\end{figure*}

\begin{figure*}
  \centering
  \includegraphics{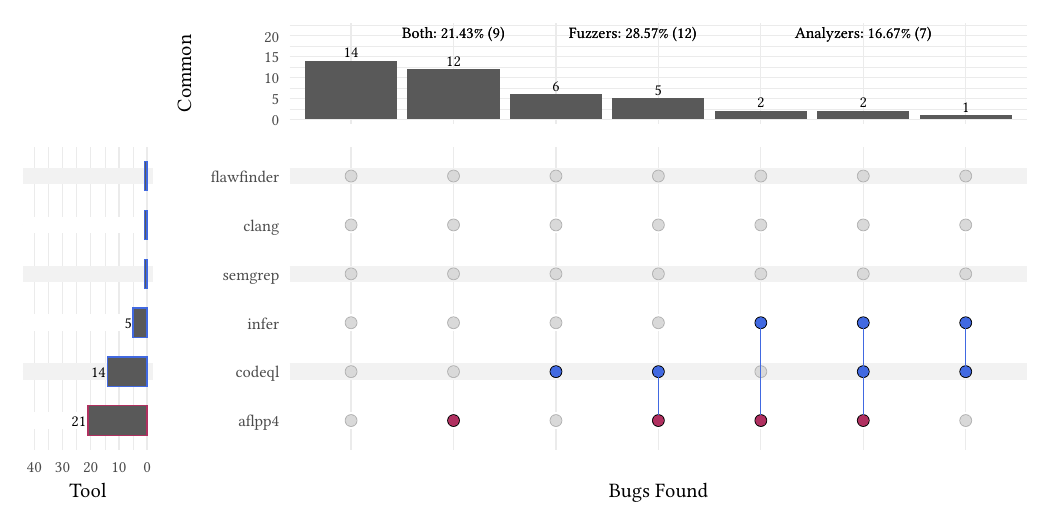}
  \vspace{-0.8cm}
  \caption{Upset plot to show overlap of bugs for \cgc.}
  \label{fig:upsetr-overall-cgc}
\end{figure*}

In this section, we present alternative versions for the figures from the quantitative evaluation (\autoref{sec:quant-eval}).
In \autoref{fig:found-totals-cgc}, we show the distribution of bugs found per bug type for \cgc, analogously to \autoref{fig:found-totals}, where we look at \magma only.
As an extension to \autoref{fig:sast-ratio-magma}, we show the number of discovered bugs in relation to the number of created reports on \cgc only in \autoref{fig:sast-ratio-cgc}, and for \cgc and \magma combined in \autoref{fig:sast-ratio-both}.
\autoref{fig:upsetr-overall} shows the bug overlap between tools for the \magma and \cgc dataset combined, analogously to \autoref{fig:upset-overall}, which looks at \magma only.
Additionally, we provide a \cgc-only version in \autoref{fig:upsetr-overall-cgc}.

\end{document}